\documentstyle[epsf,aps,prb,multicol]{revtex}
\begin{document}
\newcommand{\Pvec}{{\rm\bf P}}
\newcommand{\Evec}{{\rm\bf E}}
\newcommand{\eps}{\epsilon}  
\newcommand{\veps}{\varepsilon}  
\newcommand{\De}{$\Delta$}
\newcommand{\de}{$\delta$}
\newcommand{\mc}{\multicolumn}
\newcommand{\be}{\begin{eqnarray}}
\newcommand{\ee}{\end{eqnarray}}
\newcommand{\einf}{\varepsilon^\infty}
\newcommand{\ez}{\varepsilon^0}  

\draft \title{Non-linear macroscopic polarization in III-V nitride alloys}  
\author{Fabio Bernardini and Vincenzo Fiorentini}
\address{Istituto Nazionale per la Fisica della Materia 
and Dipartimento di  Fisica, Universit\`a di Cagliari, Cagliari, 
Italy} 
\date{\today} 
\maketitle

\begin{abstract}
We study the  dependence of macroscopic polarization on composition
and strain in wurtzite III-V nitride ternary alloys using ab initio 
density-functional techniques.  The {\it spontaneous} polarization is
characterized  by a large bowing, strongly dependent on the alloy
microscopic structure.  The bowing is due to  the   
different response of the bulk binaries to hydrostatic pressure, and to
 internal strain effects (bond alternation). Disorder effects are
instead minor. Deviations from parabolicity (simple bowing) are  of
order 10 \% in the most extreme case of AlInN alloy, much less  at
all other compositions. {\it Piezoelectric} polarization is also
strongly non-linear.  At variance with the spontaneous component, this
behavior is  independent of microscopic alloy structure or disorder
effects,   and due entirely to the non-linear strain dependence of the
{\it bulk} piezoelectric response. It is thus possible to predict the
piezoelectric polarization for  any alloy composition using the
  piezoelectricity of the parent binaries.  
\end{abstract}


\begin{multicols}{2}
\section{Introduction}
III-V nitrides and their  alloys have recently emerged as a strategic
materials system for the design and manufacture of high-frequency
light emitting devices and high power
transistors. \cite{ambacher-review} Recent theoretical investigations
\cite{noi.pol,noi.pol2,noi.mqw} have suggested that 
the macroscopic polarization of wurtzite nitrides has a major
influence on the design criteria of low-dimensional nanostructures
aimed to address the technological goals mentioned above.   
Indeed, III-V nitrides
in their wurtzite structure posses a large spontaneous 
polarization  and larger-than-usual piezoelectric
constants.\cite{noi.pol} As a consequence, fixed and approximately
two-dimensional charges resulting from polarization discontinuities 
occur at nitride heterointerfaces,\cite{noi.pol2}
 causing built-in electrostatic fields and/or free-charge accumulation,
completely unusual in 
conventional III-V's.\cite{noi.pol2,noi.mqw}
By and large, these findings have been  confirmed  experimentally
\cite{exp} through the observation of optical shifts in 
quantum wells, and of high-density two-dimensional electron gases in
field-effect transistor structures, and have been used as a guideline
in the design and optimization of non-conventional high-frequency, 
high-breakdown voltage heterostructure field-effect transistors. 

In nanostructure modeling or experiment interpretation, 
it was assumed so far \cite{noi.mqw} that polarization 
in nitride alloys would interpolate linearly  between the limiting
values determined by the parent binary compounds,\cite{noi.pol} i.e. 
follow a Vegard-like behavior. The structural properties 
 generally follow Vegard's law in  many semiconductor alloys. On  the
other hand, electronic properties such as the optical band gap
often depart from linear behavior: it is thus a plausible
expectation that macroscopic polarization may also  exhibit
non-linearity. In this paper, we investigate   the dependence of the
macroscopic polarization 
on the composition and microscopic structure in AlGaN, InGaN and AlInN
alloys, by means of ab initio density-functional techniques. Whereas
the structural properties do follow Vegard's law closely, we find a
strong non-linear dependence of the polarization on alloy composition.
The non-linearity of  the {\it spontaneous} polarization  depends
appreciably on the microscopic structure of the alloy. By contrast, the
non-linearity of  the {\it piezoelectric} polarization is a pure bulk
effect: Vegard's law holds closely if the non-linear strain dependence of
bulk piezoelectricity is accounted for.

\section{Alloy structure}

The calculation of the macroscopic polarization of an alloy requires
the knowledge of its equilibrium structure. 
Nitride alloys are commonly thought to be characterized by a random
distribution of the group-III elements on the  wurtzite cation sites,
while anion sites are always occupied by nitrogen. Recent experimental
\cite{ruterana} work (with support from theoretical \cite{neugebauer}
investigations) has suggested that ordering takes  place in some
nitride alloys  under suitable growth conditions. The  ordering in
question was reported to consist of 1$\times$1 super-lattice
structures  oriented  along the (0001) direction, with Al (or In)
atomic planes alternating with Ga planes.\cite{ruterana}      
Given these suggestions, a theoretical investigation of the alloy
phase diagram would require  the study of surface thermodynamics  and
kinetics as a function  of the growth conditions. This is outside the
scope of the present work. Here we accept as plausible that  both
random and  ordered structures may be obtained depending on the growth
conditions, and consider a sample of both structures in our
polarization calculation.  
The ordered structure just mentioned bears evident similarities with
the  CuPt-like structure found in  epitaxially ordered alloys of
zincblende materials,\cite{zunger-cupt}  where the super-lattices are
formed in 
the (111) direction; we therefore label this structure as CP.
On the other hand, the microscopic structure of  a random alloy   can
also be represented efficiently in periodic conditions   using the  
special quasi-random structure (SQS) method.\cite{SQS}  Given its
successes  in the  prediction of  band gap non-linearity in
zincblende-structure alloys, the SQS method is the natural choice to
study polarization non-linearity in III-V nitride alloys. 
We  enforce the periodic boundary conditions  needed to compute the
macroscopic polarization in the Berry phase approach \cite{KS} by
using repeated supercells, as done previously in
Ref. \onlinecite{pasquarello} to study Born charges and IR spectra in
amorphous  SiO$_2$.  As a compromise between computational workload
and the accurate description of random structures,  a 32-atom
2$\times$2$\times$2 
wurtzite supercell is adopted.  In analogy to zincblende-based
alloys,  it is possible to reproduce the statistical properties (pair
correlation functions, etc.) of a random wurtzite alloy for a molar
fraction $x$=0.5 by suitably placing the cations on the 16 sites
available in the cell, i.e. by a so-called SQS-16 structure. Even
smaller structures such as SQS-8 were previously found\cite{SQS}  to
be adequate to mimic random zincblende-based alloys, so we are
confident that our SQS-16 is a good model of the present random alloys.
Since other  molar fractions cannot be described as easily, and also
because non-linear effects are  expected to be largest for this 
concentration,  our investigation  on random alloys is restricted to
the case $x$=0.5.  

As a further ingredient, we consider ordered
structures\cite{zunger-cupt} other than
the CuPt mentioned earlier  for each of the ternaries (AlGaN, InGaN, AlInN).
The chalchopyrite-like structure (henceforth labeled  CH) is defined by each
anion site being surrounded by two cations of one species and two of
the other, with the constraint  of fitting periodically into our
(2$\times$2$\times$2) wurtzite supercell.  This structure is highly
symmetric, as  there are  only two kinds of  inequivalent  anion
sites,  differing in the orientation of the neighbors, and not in
their  chemical identity. Among the possible ordered structure, CH  is
in a sense the most homogeneous for the given composition. A further
 useful ordered structure we consider is a luzonite-like structure,
resembling the  luzonite structure\cite{zunger-cupt} used  for
zincblende-based alloys. In this  structure (labeled LZ),  each
nitrogen atoms is surrounded by 
three cations of one species, and one of the other: in a sense,
this  is  the analog to the CH structure for molar fractions $x=0.25$
and  $0.75$. 

The comparison between chalchopyrite-like and SQS ``random''
structures will give us insights on the effect of randomness versus
ordering without the biases due to specific super-lattice ordering as
in the CuPt structure. Luzonite-like structure will provide values of
the polarization  at intermediate molar fractions.

\section{Method}
As for the technical ingredients of the calculations, we work within
the density-functional-theory pseudopotential plane-wave framework as
implemented in the  VASP code.\cite{VASP} Ultrasoft pseudopotentials
\cite{uspp} provided with VASP\cite{VASP}
 are used for all the elements involved;  Ga and In $d$
semicore states are included in the  valence.  Since we are 
interested in the effects on polarization due to (among other)
internal strains related to size mismatch of the two alloyed
materials, it is imperative  that our calculation 
reproduce relative mismatches  with the highest possible
accuracy. Comparing DFT calculations using both the LDA and GGA (in its
PW91 variant\cite{pw91}) exchange-correlation
functional, we found that GGA does a better job of reproducing 
 the  relative  mismatch between the binary constituents,
besides getting closer to the experimental structure.\cite{zor}  We
therefore used the GGA for all the present calculations. We reported
recently in some detail the properties of binary nitrides, including
polarization, as calculated in the GGA vs LDA in Refs.\onlinecite{zor}
and \onlinecite{antibech}.   

Another precaution adopted throughout this work was to use the same 
32-sites supercell for both binaries and alloys,  the same k-points
for  reciprocal space integration (a 4$\times$4$\times$4
Monkhorst-Pack mesh)  and for the Berry phase calculations, and the
same safe plane-wave cutoff of 325 eV, regardless of composition. 
Although more costly, this route reduces the systematic convergence
errors. All forces and stress components
have been converged to  zero within 0.005 eV/\AA\, and 0.01 Kbar,
respectively, for free-standing alloys. For GaN-epitaxial alloys
(Sec. \ref{section.piezo}), only
the $zz$ stress component has been zeroed.  
The polarization of the binary  compounds as obtained in the
supercell calculations is always within less than 0.001 C/m$^2$ 
of the value obtained in a  bulk calculation using a
12$\times$12$\times$12 k-points mesh. 

\section{Spontaneous polarization} 
We start our analysis comparing the values of the lattice constants
and spontaneous polarization  computed for the free standing alloys, 
with the predictions of   Vegard's law. The latter relates
the lattice constants $a$ and $c$, and the
 spontaneous polarization $P_{\rm sp}$ of an alloy A$_{\rm x}$B$_{\rm
1-x}$N, to their values for the binary costituents AN and BN 
(A and B  indicate two generic cations among In, Al, and Ga) by  
\be    
  a\,({\rm A_{\rm x}B_{\rm 1-x}N}) = x\, a({\rm AN}) + (1-x)\, a({\rm BN}),
\label{eq.a}
\ee
\be    
  c\,({\rm A_{\rm x}B_{\rm 1-x}N}) = x\, c({\rm AN}) + (1-x)\, c({\rm BN}),
\label{eq.c}
\ee
and
\be
  P_{\rm sp} ({\rm A_{\rm x}B_{\rm 1-x}N}) = x\, P_{\rm sp}({\rm AN}) + (1-x)\,
P_{\rm sp}({\rm BN}).  
\label{eq.pol}
\ee     
The  existence of compositional non-linearities
(i.e. non-Vegard behavior)  in a generic measurable quantity in an
alloy can be described in a first approximation by a parabolic  model 
involving a so-called {\it bowing} parameter. In particular, we
expect to be able to describe the possible non-linear dependence of
the spontaneous polarization   on the composition as
\be
P ({\rm A_x B_{1-x} N}) &=& x\, P({\rm AN}) + (1-x)\, P({\rm BN}) \nonumber\\
                     &\ & -\,  b_{\rm AB}\, x\,(1-x),
\label{eq.bow}
\ee
where the bowing parameter  is  by definition
as
\be
   b_{\rm AB} = 2 P({\rm AN}) + 2 P({\rm BN})
-4 P({\rm A}_{0.50}{\rm B}_{0.50}{\rm N})
\ee
involving only the knowledge of the polarization for ternary alloys
with molar fraction $x=0.5$. Given the bowing parameter, the 
 spontaneous polarization can be obtained at any composition.

\subsection{Polarization non-linearity}

\begin{figure}[h]
\epsfclipon
\epsfxsize=8cm
\epsffile{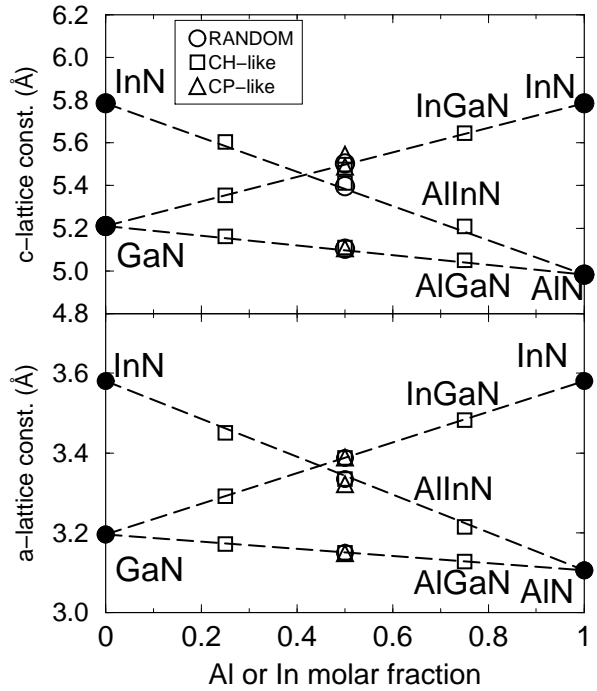}
\caption{Basal ($a$) and axial ($c$)  lattice constants of
wurtzite-based nitride alloys. Directly calculated values for various
alloy structures  are denoted by open circles (random), squares (CH
and LZ), triangles (CP). The dashed lines are Vegard's law.}
\label{fig.ac}
\end{figure}

\begin{figure}
\epsfclipon
\epsfxsize=8.2cm
\epsffile{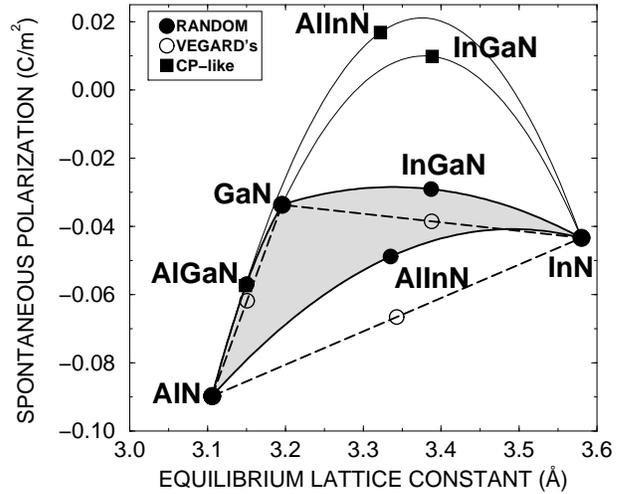}
\caption{Spontaneous polarization versus equilibrium lattice
constant in free-standing random and CuPt-ordered
nitride alloys (filled symbols).
Solid lines are interpolations according to Eq. \protect\ref{eq.bow}.
Dashed lines and open symbols are Vegard predictions.}
\label{fig.sp}
\end{figure}

\begin{figure}
\epsfclipon
\epsfxsize=8cm
\epsffile{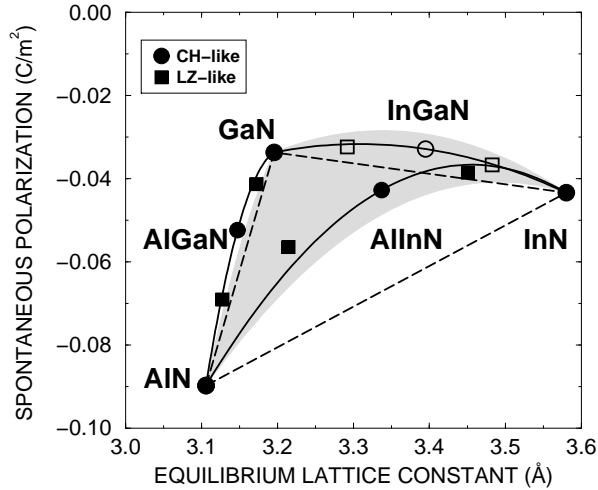}
\caption{Spontaneous polarization versus equilibrium lattice
constant for  free-standing chalcopyrite- and luzonite-like ordered
nitride alloys (filled symbols; open symbols are used for InGaN
for clarity).
Solid lines are interpolations according to Eq. \protect\ref{eq.bow}.
Dashed lines are Vegard predictions.
The shaded triangle represents the polarization for the random
alloys as reported in Fig. \protect\ref{fig.sp}.}
\label{fig.sp2}
\end{figure}

In Fig. \ref{fig.ac}, we compare  with Vegard-law predictions our
calculated equilibrium lattice parameters $a$ and $c$ for all the
alloy structures and compositions studied. Both $a$ and $c$ follow
quite closely Vegard's law in all cases. The dependence of the
polarization on composition is then the same as that on  the 
lattice  parameter(s), modulo a multiplicative factor. Therefore, in
Figs.\ref{fig.sp} and \ref{fig.sp2} we can compare with Vegard-law
predictions the calculated  data, respectively,  for the random and
Cu-Pt--ordered (Fig.\ref{fig.sp}), and chalcopyrite- or luzonite-like
ordered  (Fig. \ref{fig.sp2})  alloys, displayed in the (P,$a$)
plane.  The ordinates of the filled symbols are the calculated
spontaneous polarization, the abscissas are the calculated lattice
constants; the open symbols and dashed lines are the Vegard-law
predictions; the solid lines are fits of Eq.\ref{eq.bow} to the
calculated values.

We identify two main  points. {\it First}, there is a large non-linear
dependence of the spontaneous polarization on the lattice parameter,
hence on alloy composition. Alloys with large  lattice mismatch
between the constituents (e.g. AlInN and InGaN) show the largest
bowing, while in AlGaN the non-linearity is modest, although
sizable. Regardless of the alloy structure and composition, the
spontaneous polarization shows always an upward bowing. {\it Second},
CP-ordered (Fig. \ref{fig.sp}) and CH-ordered (Fig. \ref{fig.sp2})
alloys have very different bowing.  This is at variance to the case of
the band gap: ordering is known to  cause an increase of the band gap
bowing, but more homogeneous across different ordered structures, and
never  as  dramatic as that of polarization in the CP ordered alloy.  

\vspace*{-1.0cm}
The CH and LZ ordered structures offer the possibility to test the 
validity of the interpolation model based on the bowing parameter.
The ordered LZ structure is  analogous to CH for molar fractions of
0.25 and 0.75. The size of the  deviations of the LZ values from those
predicted by the fit of   Eq.\ref{eq.bow} through  the binaries and
the CH ($x$=0.5) value,  indicates whether or not non-parabolicity
occurs in the $P$($x$)  relation for CH-like order. Since the
polarization of the CH structure  behaves  qualitatively as that of
the  random structure (Fig. \ref{fig.sp2}),  the  conclusions for CH
may be exported to the random phase. 
In Fig.~\ref{fig.sp2} the values of the polarization calculated for
the LZ structures at molar fractions  0.25 and 0.75 lay very close to
the parabolic  curve  for InGaN and  AlGaN; this confirms the accuracy
of the quadratic hypothesis embodied in Eq.\ref{eq.bow}. For  AlInN,
the   calculated values  lay somewhat above  (below) the parabolic
relation  for $x$=0.25  ($x$=0.75), showing  that some polarization
non-parabolicity occurs, and specifically  that the bowing is higher
for low In concentration in  AlInN.  Compared to the quadratic
non-linearity (Eq.\ref{eq.bow}),  the non-parabolicity 
causes  relatively modest   additional deviations (of order 10 \%)
for AlInN. A legitimate conclusion is then that the
bowing relation should  predict polarization in AlInN with about 
10 \% accuracy, and much  better for InGaN and AlGaN.

It is clear that these result have a considerable technological
relevance for nanostructure devices. Polarization differences
across  interfaces in, say, multi-quantum-wells, generate localized
interface charges \cite{noi.pol} which in turn cause electrostatic
fields and/or charge accumulation at the interfaces.\cite{noi.mqw,exp}
The non-linearity just reported modifies  considerably the
polarization-induced interface charge  densities, and must be
accounted for in any attempt at quantitative  comparisons with, or
predictions on experiments. We will be considering this aspect in
detail in forthcoming work. In the following Sec.\ref{subsec.origin} 
we attempt instead to gain insight on
the  nature  of the non-linear behavior of the polarization in III-V
nitrides. 

\subsection{Microscopic origin of polarization non-linearity}
\label{subsec.origin}

 It is known from earlier theoretical work on III-V zincblende-based
 alloys \cite{Jaffe} that band gap non-linearities can be attributed
 to essentially three sources:
{\it i)} chemical effects due to the different cation electronegativity of
 the components; {\it ii)} internal strain effects, due to 
 varying cation-anion bond lengths (the so-called bond alternation);
{\it  iii)} disorder effects due to the random distribution of the
chemical elements on the cation sites.
Chemical effects and internal strain are dominant,  whereas disorder
 gives only a modest contribution. On the basis of these indications,
 one expects a large bowing in  random alloys  whose constituents have
 a large lattice mismatch. Such a behavior for the fundamental gap
 bowing was  indeed found  for zincblende III-V nitrides
 alloys\cite{Wright}.  Also, the bowing  in  ordered  structures
 should be similar to, but larger than in  random  structures.  

For the polarization,  we only find a qualitative behavior similar to
the band gap bowing  in the case of  random  alloys, while the
ordered phases CH and CP do not  seem  to show identifiable trends. It
is especially surprising that i) the random and CP phases of AlGaN have
nearly the same bowing; ii) the InGaN CH ordered alloy  shows a {\it
smaller} bowing than its random phase. 
The reason for this anomalous behavior must be found in the peculiar nature
of the polarization bowing. While the band gap is a scalar,
the polarization is a vector of fixed direction, namely, for
wurtzites, the (0001) direction. Thus, bond alternation will affect 
the polarization bowing only if it  changes the projection of
the bond length in the (0001) direction.
This idea is supported by the fact that in pure binaries the
polarization is strongly influenced  by the relative displacement of
the cation and anion sub-lattices in the (0001)
direction.\cite{antibech} 

\begin{figure}
\epsfclipon
\epsfxsize=8cm
\epsffile{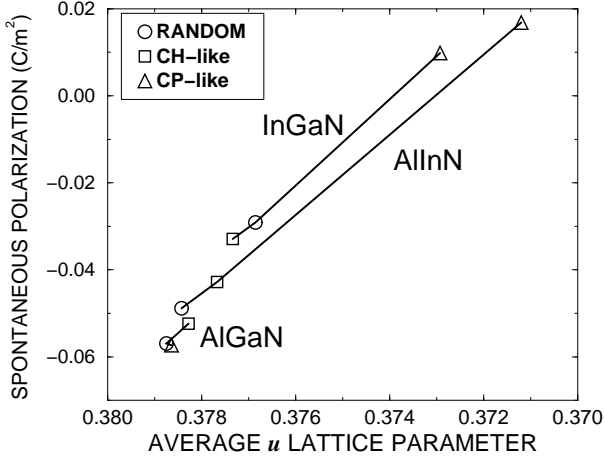}
\caption{Spontaneous polarization versus  average {\it u}
parameter in ternary alloys. Open circles, squares, triangles,
 refer to random, CH-like, CP-like structures, respectively.  }
\label{fig.pu}
\end{figure}

 Indeed, we find that a clear correlation exists between the so-called
{\it u}  parameter of the wurtzite structure, i.e. the bond length
along the singular polar (or pyroelectric) axis, and the value of the
polarization. In Fig. \ref{fig.pu}, we show the calculated spontaneous  
polarization of free-standing binaries and $x$=0.5--alloys
 in the various structures, vs. the
{\it average} internal parameter {\it  u}. The  latter is defined as
the average value of the projection  of the  connecting vector of a
nitrogen atom with its   first neighbor in the
(000$\bar{1}$) direction along this same direction.\cite{nota}
This  definition can be used also for
random phase alloys in spite of the displacements off the ideal sites. 
We see in Fig. \ref{fig.pu} that for a given alloy composition, the
 spontaneous polarization of free-standing alloys of different microscopic
structure depends   linearly on the average {\it u} parameter of 
the alloy structure.   This indicates that spontaneous polarization
differences between alloys  of the {\it same composition} are due
essentially to structural and bond alternation effects;   disorder
appears to  have a negligible influence.   
It appear instead that the chemical identity of the constituents plays
a role of some importance (beyond the obvious fact that it implicitly
determines the structure).

The  points mentioned above are now easily explained. The random
and CP phases in AlGaN have almost  the same average $u$, hence almost
the same polarization. In InGaN, the random alloy has a larger $u$
than the CH phase, while the opposite holds for AlInN and AlGaN:
therefore the  CH-vs-random bowing behavior in InGaN is opposite with
respect to the others. Also, the huge bowing of  CP-ordered AlInN
and InGaN matches the very large deviation of the average $u$ in those
phases from the random and CH-like structures.
If structure were the only source of polarization bowing,
all of the points in Fig. \ref{fig.pu} would fall on the same straight line. 

To check this role, we set up a model based
on the polarization  of binaries and  alloys in a  constrained  {\it 
 ideal} wurtzite structure. In this structure, only the $a$ parameter
is independent, whereas $c$ and $u$ are fixed at the values determined
by maximal sphere packing, namely  {\it u}=0.375 and {\it
c/a}=$\sqrt{8/3}$. Each nitrogen atom  is then surrounded by four
equidistant cations -- that is,  all  bonds have  the same length for
a  given lattice constant $a$. By construction, then,  bond alternation
do not play any role, and the  effects of chemical identity
of the constituents can be easily disentangled.  

We assume that Vegard's law holds for the lattice constant $a$,
as was found to be the case  for the alloys in the preceeding
Subsection.
This establishes a (linear) relation between composition and lattice
constant. We then calculate the polarization in each of the {\it
binary} nitrides in their ideal structure as  function of the lattice
constant $a (x)$. Finally, we express the alloy polarization as  a 
composition-weighted Vegard-like average of the  polarizations of the
binaries as 
\be
P_{sp}({\rm A}_x{\rm B}_{1-x}{\rm N}) = x P_{sp}^{a(x)}({\rm AN})  
                       + (1-x) P_{sp}^{a(x)}({\rm BN}). 
\label{eq.interp}
\ee
In this model, any non-linearity must come from the different 
response of polarization to changes in $a(x)$, hence to
 hydrostatic compression, in the ideal binary compounds.

\begin{figure}
\epsfclipon
\epsfxsize=8cm
\epsffile{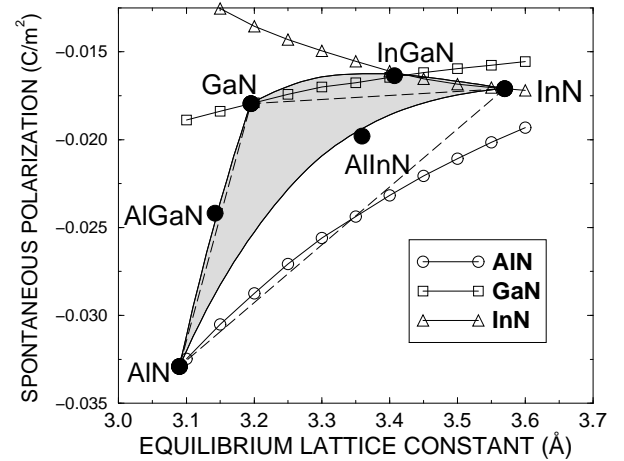}
\caption{Spontaneous polarization versus lattice
constant for the  {\it ideal wurtzite} structure.
Filled circles refer to the values of binary compounds and random ternary
alloys. Open circles, squares and triangles are the polarizations
calculated as function of the lattice constant in bulk AlN, GaN, and InN.
Solid lines are Vegard interpolations of the latter values.
This shows that the effect of
 hydrostatic  pressure on bulk polarizations reflects directly on
the polarization in the alloys.}
\label{fig.spid}
\end{figure}

We report in Fig. \ref{fig.spid} three sets of quantities: the
polarizations computed in the {\it ideal} wurtzite structure for the
three ternary alloys   at their calculated equilibrium volume (filled
circles);  the polarization of the  binaries in the ideal structure as
a function of the lattice parameter {\it a} (open symbols); the Vegard
interpolation Eq.\ref{eq.interp} of the  latter polarizations  (the
bowed shaded triangle) for the alloys.  We see that the computed
values for the alloys and the Vegard prediction  essentially
coincide. Therefore, the origin of the ``chemical'' non-linearity, 
and	its large values in In-containing alloys,  becomes clear:
in AlN and GaN, the polarization decreases with hydrostatic pressure,  
while it increases in InN.  Also to be noted,  polarizations in the
ideal structure are between  35 to 50 \% of their value in free
standing alloys,  and despite the absence of bond alternation, the
bowing is still very large.

The above model provides an expression of the  bowing parameter
of {\it ideal} wurtzite structure alloys as function of the polarization  
response to hydrostatic pressure:
\be
b^{\rm AB}_{\rm model} 
&=&  (a_{\rm BN} - a_{\rm AN}) 
\left( \frac{\partial P_{\rm BN}} {\partial a} - 
       \frac{\partial P_{\rm AN}} {\partial a}  \right)
                   \vert_{a=a(1/2)}\\ 
&+&\frac{1}{4} (a_{\rm BN} - a_{\rm AN})^2
\left( \frac{\partial^2 P_{\rm AN}} {\partial a^2} +
       \frac{\partial^2 P_{\rm BN}} {\partial a^2}  \right)
                   \vert_{a=a(1/2)}\ .\nonumber
\ee
The agreement of the latter expression with the $b$ resulting
from a  fit to the calculated values is very good (e.g. for the extreme
case of AlInN,  $b_{\rm model}$=--0.0225  C/m$^2$, while
from direct calculation we get --0.0208 C/m/$^2$).
On the basis of the model, it is now understandable that the AlGaN bowing
is pretty moderate since  the region of interest is small
 (3.1 \AA\, to 3.2 \AA) and the 
responses to  hydrostatic pressure of AlN and GaN are similar.
On the contrary, in the large range  3.1--3.6 \AA, AlN and InN have
opposite  behaviors, whence  the huge bowing found in AlInN
alloys. The same goes, although to a lesser extent, for InGaN alloys.

\section{Piezoelectric polarization}
\label{section.piezo}

The knowledge of the spontaneous polarization in alloys is not 
sufficient to predict the values of interface charge accumulations
or  electrostatic fields in wurtzite-nitride--based
nanostructures, because the latter are usually grown
pseudomorphically and under strain on the (0001)
or (000$\bar{1}$) faces of a substrate (typically sapphire or SiC, 
with or without AlN or GaN buffer layers). The ensuing
symmetry-conserving strain causes a change in polarization that amounts
to a {\it piezoelectric} polarization. In general, the polarization in
the active layers of a nanostructure  is a 
combination of spontaneous  and piezoelectric polarization.   
In this context, the aim of this Section is threefold. First, we
demonstrate that  piezoelectricity in nitride alloys is non-linear;   
second, we show that this non-linearity is due to
a pure bulk effect, i.e. the non-linear behavior of
bulk binary piezoelectric constants vs symmetry-conserving strain; third,
we suggest how to use this understanding in practice to predict
piezoelectricity in any nitride alloy.

We consider for definiteness (but with no loss of generality as to the
scheme to predict piezoelectricity in arbitrary alloys)
the technologically relevant case of an
alloy pseudomorphically grown on an unstrained GaN substrate.
Using the same technical ingredients as previously, we repeat the 
polarization calculations with the costraint that 
$a^{\rm alloy}$=$a^{\rm GaN}$, fully reoptimizing all structures. 
We compute the piezo component as the difference of  the 
total polarization obtained in this calculation and the 
spontaneous part calculated previously for
the free-standing alloy.  

\begin{figure}
\epsfclipon
\epsfxsize=8cm
\epsffile{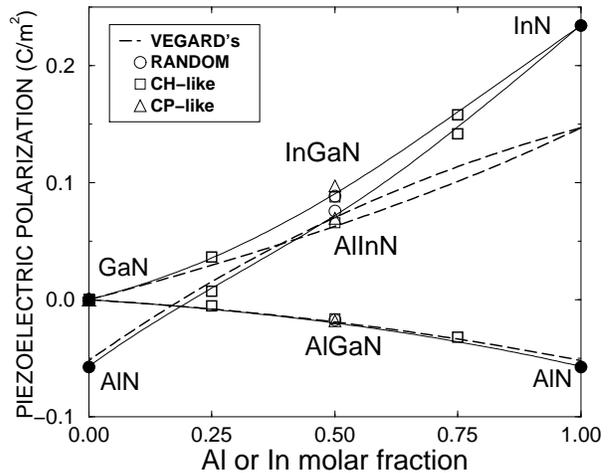}
\caption{Piezoelectric component of the macroscopic
polarization in ternary alloys epitaxially strained on a GaN substrate.
Open symbols are directly calculated values for
random (circles), CH-like and LZ-like (squares), and
 CP-like (triangles)  structures, respectively.
Dashed lines are the prediction of linear piezoelectricity, while the
solid lines is the prediction of Eq.\ref{eq.pz}
using the non-linear bulk polarization as reported in
Fig. \protect\ref{fig.non-lin-pz}.}
\label{fig.pz}
\end{figure}

In Fig. \ref{fig.pz} we show the piezoelectric  polarization as a
function of the alloy composition.  Symbols are the calculated
polarizations in the various alloy structures and compositions; it is
clear that, contrary to  spontaneous part, the piezoelectric
polarization component hardly depends on the microscopic structure of
the alloy.  We now ask whether the piezoelectric polarization of the
alloy can be reproduced by a Vegard-like model containing only the
knowledge of the properties of the binaries, i.e.
\be
   P^{\rm ABN}_{\rm pz}(x) = 
                x P_{\rm pz}^{\rm AN}  +  (1-x) P_{\rm pz}^{\rm BN}.
\label{eq.pz}
\ee
In a first approximation, one may  calculate the piezoelectric
polarization of the binary compounds for symmetry-conserving  in-plane
and axial strains as 
\be
 P_{\rm pz}^{\rm AN} = e_{33}\, \eps_3 + 2\, e_{33}\, \eps_1, 
\label{eq.pz2}
\ee
where the piezoelectric constants $e$ are calculated\cite{noi.pol} 
in the equilibrium structure of the binary AN, and by definition {\it
do not depend} on strain. 
The dashed lines in Fig. \ref{fig.pz} represent the piezoelectric
term  as computed from  the above relations using the 
piezoelectric  constants computed \cite{noi.pol} for the binaries.
Clearly, {\it when   combined with Eq.\ref{eq.pz2}}, Vegard's law
(Eq.\ref{eq.pz}) fails to reproduce the  calculated polarization, and
misses the  strong non-linearity of the 
piezoelectric term  evident in Fig. \ref{fig.pz}.

\vspace*{-0.5cm}

\begin{figure}
\epsfclipon
\epsfxsize=7.6cm
\epsffile{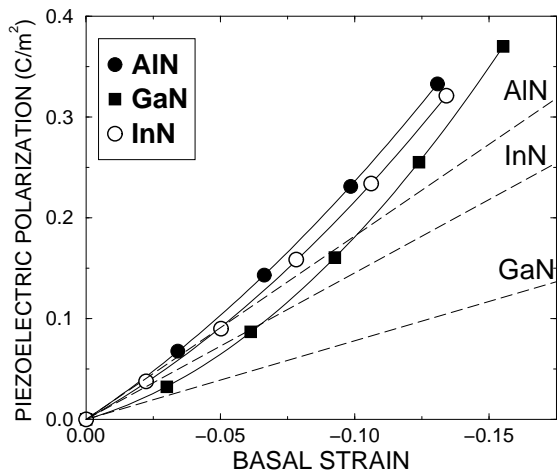}
\caption{Piezoelectric
polarization in binary nitrides as function of basal strain
(symbols and solid lines) compared to linear piezoelectricity
prediction (dashed lines). The $c$ and $u$ lattice parameters are optimized
for
each
strain.}
\label{fig.non-lin-pz}
\end{figure}

This is due to a genuine non-linearity of the bulk piezoelectricity of
the binary constituents, of non-structural origin; hence, the bowing
due to the microscopic structure of the alloys is negligible. The
proof of this statement is in two steps. First, we  directly calculate
the  piezoelectric polarization  as a function  of basal strain 
for  AlN, GaN, and InN, optimizing all structural parameters. The
results, collected in Fig. \ref{fig.non-lin-pz}, clearly
 show that the piezoelectric polarization of the {\it binaries} is
an appreciably non-linear function of $a$, i.e. of basal strain.
Since all lattice parameters closely follow Vegard's law, the
non-linearity cannot be related to deviations from linearity in the
structure.  

Second, we plug   the  {\it non-linear} piezoelectric
 polarization  just computed for the binaries into the  Vegard
interpolation Eq.\ref{eq.pz}. We obtain thereby the
solid lines in Fig. \ref{fig.pz}:  we see that the predictions
of this scheme are in excellent agreement with the polarization
calculated directly for the alloys. Therefore, we can conclude that 
{\it a)} the   non-linearity in bulk piezoelectricity dominates over
 any effects related to disorder, structure, bond alternation, etc.,
 and  that 
{\it b)} Vegard's law holds for the piezoelectric polarization of
 III-V nitrides alloys, {\it  provided that the non-linearity of the bulk
 piezoelectric  of  the constituents is accounted for}. 

The latter conclusion suggests a straightforward scheme for predicting
the piezoelectric polarization of any nitride alloy at any strain:
pick a value for $x$, calculate the basal strain $\veps(x)$ from
Vegard's law, calculate $P_{\rm pz}$ from Eq.\ref{eq.pz} using the
non-linear 
piezoelectric polarization of the binaries
(Fig. \ref{fig.non-lin-pz}). This scheme is of obvious interest in
the modeling of nanostructures, especially for high In contents and
AlInN alloys.

\section{Summary and acknowledgements}

In this paper, we studied the non-linear dependence of the polarization 
on composition and strain in ternary III-V nitride wurtzite-based alloys.
The spontaneous polarization is  characterized by a large bowing,
strongly dependent on the alloy microscopic structure. The bowing is
due to the different response of the bulk binaries to hydrostatic
pressure, and to internal strain effects due to bond
alternation. Disorder effects are minor. Deviations from parabolicity
(simple bowing) are  of order 10 \% in the most extreme case (the
AlInN alloy), much less for all other compositions. 

Piezoelectric polarization is also strongly non-linear. This behavior
is  independent of  microscopic structure or disorder effects
(the structural parameters for all alloys closely follow Vegard's law and
linear elasticity), and is due entirely to the non-linear strain
dependence of the  bulk piezoelectric response. Based on our
understanding of this behavior, we suggested that it is possible to
predict the (non-linear) piezoelectric polarization for  any alloy
composition using the knowledge of the binary compound
piezoelectricity.   

We acknowledge support from a MURST Cofin99 project and from 
Iniziativa Trasversale Calcolo Parallelo of INFM.

\end{multicols}

\begin{thebibliography}{99}
\bibitem{nota}
We adopt for the  wurtzite structure  the convention according to which
each anion sits at (0,0,u) from the cations, i.e. all the vertical bonds
between, say, Ga and N point along (0001).
\end{thebibliography}
\end{document}